\documentclass{aa}
\usepackage{graphics}
\begin{document}

\def\frac{$''$\hspace*{-.1cm}}
\def\min{$'$}
\def\deg{$^{\circ}$\hspace*{-.1cm}}
\def\min{$'$\hspace*{-.1cm}}
\def\h2{H\,{\sc ii}}
\def\hi{H\,{\sc i}}
\def\hb{H$\beta$}
\def\ha{H$\alpha$}
\def\oiii{[O\,{\sc iii}]}
\def\sm{$M_{\odot}$}
\def\ab{$\sim$}
\def\x{$\times$}
\def\sec{s$^{-1}$}

\title{Resolving the compact \h2 regions in N160A with {\it HST}
\thanks{Based on observations with the NASA/ESA Hubble Space
Telescope obtained at the Space Telescope Science Institute, which is
operated by the Association of Universities for Research in Astronomy,
Inc., under NASA contract NAS\,5-26555. These observations are
associated with proposal \#8247.}}

\offprints{M. Heydari-Malayeri, heydari@obspm.fr}

\date{Received 24 September 2001 / Accepted 29 October 2001}

\titlerunning{LMC N160A}
\authorrunning{Heydari-Malayeri et al.}

\author{M. Heydari-Malayeri\inst{1} 
	\and 
  	V. Charmandaris \inst{2}
	\and 
  	L. Deharveng \inst{3}
	\and
        F. Meynadier \inst{1}
        \and
	M.\,R. Rosa \inst{4,}\,\thanks{
    Affiliated to the Astrophysics Division, Space Science Department of
    the European Space Agency.}
	\and   
	D. Schaerer \inst{5} 
	\and  
    	H.  Zinnecker \inst{6}  }

\institute{{\sc demirm}, Observatoire de Paris, 61 Avenue de l'Observatoire, 
F-75014 Paris, France 
\and 
Cornell University, Astronomy Department, 106 Space Sciences Bldg.,
Ithaca, NY 14853, USA
\and
Observatoire de Marseille, 2 Place Le Verrier, F-13248 Marseille Cedex
4, France
\and
Space Telescope European Coordinating Facility, European Southern
Observatory, Karl-Schwarzschild-Strasse-2, D-85748 Garching bei
M\"unchen, Germany
\and 
Observatoire Midi-Pyr\'en\'ees, 14, Avenue E. Belin, F-31400 Toulouse,
France
\and 
Astrophysikalisches Institut Potsdam, An der Sternwarte 16, D-14482
Potsdam, Germany}

\abstract{Using high-resolution imaging with the {\it Hubble Space 
Telescope}, we study the Large Magellanic Cloud \h2 region N160A 
and uncover several striking features of this complex massive
star-forming site.  The two compact high excitation \h2\, blobs (HEBs)
A1 and A2 are for the first time resolved and their stellar content
and morphology  is revealed. A1, being of higher excitation, is
powered by a single massive star whose strong wind has created a
surrounding bubble. A2 harbors several exciting stars enshrouded 
inside large quantities of dust.  The whole N160A nebula is energized
by three star clusters for which we obtain photometry and study their
color-magnitude diagram. The \h2 region is particularly dusty, with
extinction values reaching an $A_{V}$\,\ab\,2.5 mag in the visible,
and it is separated from the molecular cloud by an outstanding
ionization front. A previously detected infrared young stellar object
is also accurately located with respect to the \h2 region.
\keywords{
	Stars: early-type -- dust, extinction -- \h2\, regions --
	individual objects: N160A -- Galaxies: Magellanic Clouds } }

\maketitle

\section{Introduction}

The region of 30 Doradus has long been identified as a unique star
formation site not only in the Large Magellanic Cloud (LMC) but also
in  our whole Local Group of galaxies. As a result, the regions
lying at its periphery have also attracted special attention over the
years. In particular, the chain of bright \h2 regions lying south of
30 Doradus, i.e. N158, N159, and N160 (Henize \cite{hen}), not only
show several signs of ongoing star formation, such as  stars still
embedded in their birth cocoons, infrared sources, and masers, but
they are also associated with the most important concentration of
molecular gas in the LMC (Johannson et al. \cite{joh} and references
therein).\\

This paper is devoted to N160A (NGC\,2080), the brightest component of
the N160 complex, also known as DEM\,284 (Davis et al.\cite{dem}) or
MC\,76 (McGee \& Milton \cite{mc}), which lies some 30\min\, (450 pc)
south of 30 Dor.  N160A is a particularly interesting region of star
formation as it harbors several compact IR sources (Epchtein et
al. \cite{epchtein}, Jones et al. \cite{jon}) and OH and H$_{2}$O
masers (Caswell \& Haynes \cite{caswell1}, Gardener \& Whiteoak
\cite{gardener}; Caswell \cite{caswell2}, Whiteoak et al. \cite{w83},
Whiteoak \& Gardener \cite{w86}, Brooks \& Whiteoak \cite{brooks}). It
also contains the OB association LH\,103 which embodies 41 blue stars
(Lucke \& Hodge \cite{lh}, Lucke \cite{lucke}). \\

The first detailed study of N160A was carried out by Heydari-Malayeri
\& Testor (\cite{hey86}, hereafter Paper I) using extensive
optical imaging and spectroscopy of both its gaseous and stellar
content, as well as high resolution radio continuum mapping at 843
MHz.  They discovered two compact \h2\, regions embedded in the bright
\h2\, region N160A. Identified as N160A1 and N160A2, these objects
belong to the special class of so-called High Excitation Blobs (HEBs)
in the Magellanic Clouds.  In contrast to the typical \h2\, regions of
the Magellanic Clouds, which are extended structures spanning several
minutes of arc on the sky and powered by a large number of hot stars,
HEBs are very dense small regions usually 5\frac\, to 10\frac\, in
diameter.  At the distance of the Magellanic Clouds this corresponds
to sizes of more than 50 pc for normal \h2\, regions and 1 to 3 pc for
the blobs. HEBs are in fact associated with young massive stars just
leaving their parent molecular cloud (see Heydari-Malayeri et
al. \cite{hey01} for references).  Using extensive near-IR
observations, Jones et al. (\cite{jon}) studied N160A, obtained the
first $J$ and $K$ band images of this region, and confirmed the high
extinction of both blobs A1 and A2, which they called objects \#5 and
\#9 respectively. More recently, Henning et al. (\cite{henning}) using
near- and mid-IR images and ISO-SWS spectra established the position
of the young stellar object in N160A discovered by Epchtein et
al. (\cite{epchtein}).\\

In this paper we use observations obtained with the {\it Hubble Space
Telescope} to study the \h2 region N160A.  The higher resolution of
{\it HST} is essential in order to reveal the various emission and
dust features of the nebula on the whole and to study the so far
elusive HEBs A1 and A2. It is also necessary for better understanding
of massive star formation in this interesting region, i.e. to unveil
its stellar content and identify the exciting stars, which up to now
have remained unknown. \\

\section{Observations and data reduction}

The observations of N160A were performed with the Wide Field Planetary
Camera 2 (WFPC2) on board of the {\it HST} using several broad- and
narrow-band filters. The images taken with the broad-band filters
(F300W, F467M, F410M, and F547M) were obtained on February 5, 2000 and
aimed at revealing the details of the stellar content of N160A which
was centered on the Planetary Camera (PC). The narrow-band filter
images (F487N, F503N and F656N) were obtained on May 28, 2000. In that
case the target was centered on the WF2 which has larger pixels and
lower noise than the PC CCD and is better suited for detecting faint
nebular emission. Exposures were taken at different pointings
dithered by 0\frac.8 to better sample the point spread function,
while the exposure times ranged from 10 to 300 sec (see
Table\,\ref{obs} for details). \\

\begin{table}[!h]  
\caption[ ]{Observations of N160A ({\it HST} GO-8247)} 
\label{obs} 
\begin{flushleft}  
\begin{tabular}{lcr}  
\hline 
{\it HST} filter & Wavelength & Exposure time\\ & $\lambda$(\AA) &
(sec)\\
\hline 
F300W (wide-U)                  & 2911          & 8\,\x\,14\,=\,112\\
F410M (Str\"{o}mgren $v$)       & 4090          & 8\,\x\,50\,=\,400\\
F467M (Str\"{o}mgren $b$)       & 4669          & 8\,\x\,35\,=\,240\\
F547M (Str\"{o}mgren $y$)       & 5479          & 8\,\x\,10\,=\,80\\
F487N (H$\beta$)                & 4866          & 4\,\x\,260\,=\,1040\\
F502N (\oiii)                  & 5013          & 4\,\x\,300\,=\,1200\\
F656N  (H$\alpha$)              & 6563          & 4\,\x\,260\,=\,1040\\
\hline    
\end{tabular} 
\end{flushleft}   
\end{table}

The data were processed through the standard {\it HST} pipeline
calibration.  Multiple images were co-added using the {\sc stsdas}
task {\it imcombine}, while cosmic rays were detected and removed with
the {\sc stsdas} task {\it crrej}.  Normalized images were then
created using the total exposure times for each filter.  To extract
the positions of the stars, the routine {\it daofind} was applied to
the images by setting the detection threshold to 5$\sigma$ above the
local background level.  The photometry was performed setting a
circular aperture of 3--4 pixels in radius in the {\it daophot}
package in {\sc stsdas}. \\

\begin{figure*}[!ht]
\begin{center}
\resizebox{17cm}{!}{\includegraphics{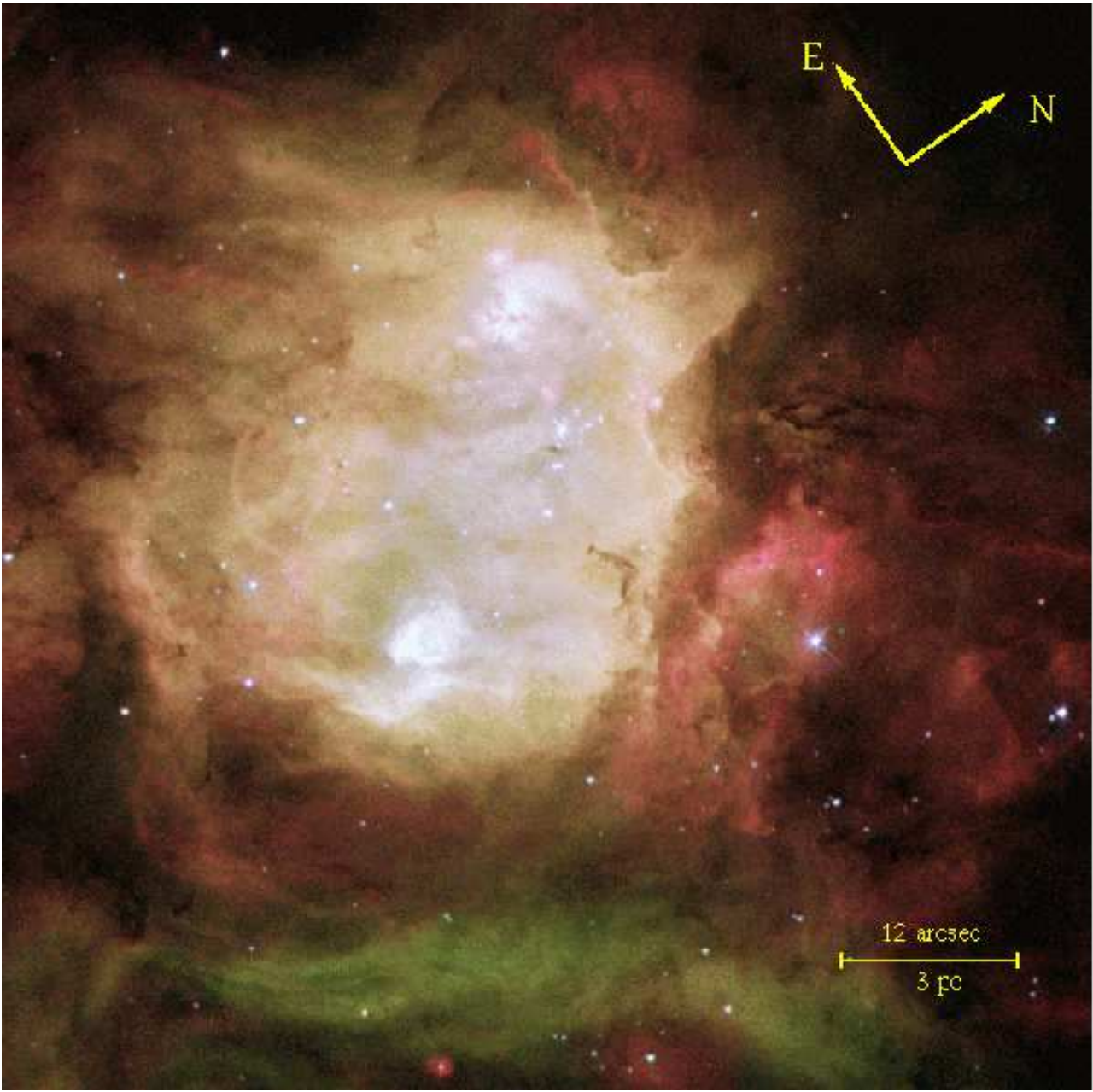}}
\caption{A ``true color'' composite image of the LMC 
\h2\, region N160A as seen by {\it HST}/WFPC2, based on images taken 
with filters \ha\, (red), \oiii\, (green), and \hb\, (blue).  The two
bright emission nebulae are the high excitation \h2 blobs A1 (bottom),
and A2 (top).  Note the cavity of a mere 2\frac.3 across (\ab\,0.6 pc)
and the surrounding shell carved by a massive star inside A1 (\#22),
as well as the bright neighboring ridge. A2 is resolved into a mottled
structure due to high dust content and harbors several faint stars.  A
prominent ionization front borders a high absorption zone to the west
where the extinction approaches values higher than $A_{V}$\,=\,2.5 mag
in the visible. There are also several conspicuous arcs and filaments
created by the winds and shocks of the embedded stars as well as a few
``tiny'' dust pillars, along with a blue star cluster towards the
middle of N160A.  The whole nebula resembles the features of the
head from a ``ghost'' or ``beast'' with the HEBs as its eyes and the
absorption zone as its mouth!  The field size is
\ab\,63\frac\,\x\,63\frac\, (\ab\,16\,pc\,\x\,16\,pc). }
\label{true}
\end{center}
\end{figure*}

\begin{figure*}
\begin{center}
\resizebox{13cm}{!}{\includegraphics{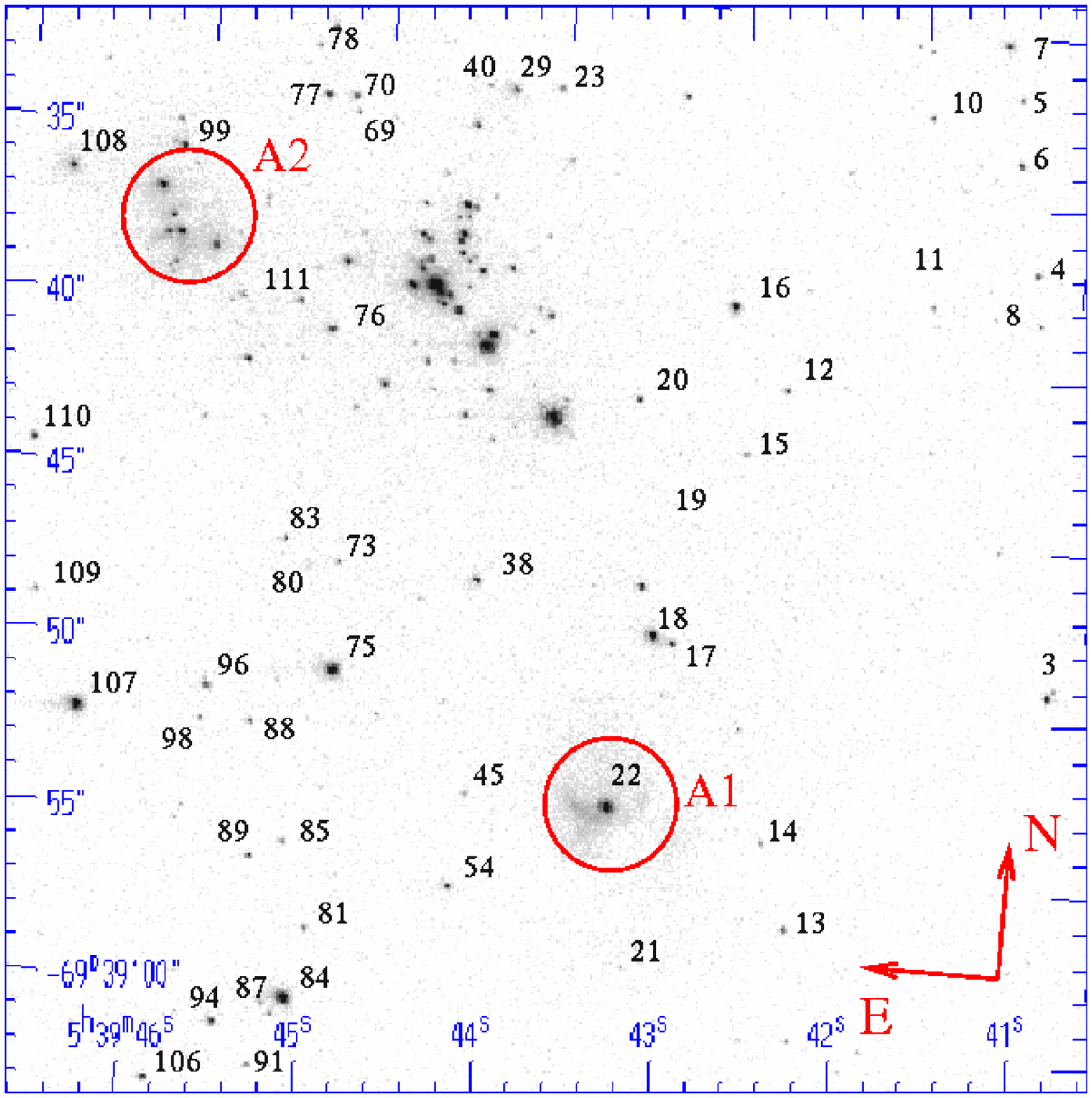}}
\caption{An {\it HST}/WFPC2 image of LMC N160A taken with the Str\"omgren 
$y$ filter (F547M) showing the stellar content of the giant \h2\,
region.   Star \#22 is associated with A1, while stars \#108, \#29,
and \#72 are identified with the three distinct red sources in the
upper center area of Fig.\,\ref{true}. The locations of A1 and A2 are
marked and the unlabeled stars are identified in the two subsamples
displayed in Fig.~\ref{clusters} and the photometry is presented in
Table\,\ref{phot}.  The field size is \ab\,32\frac\,\x\,32\frac\,
(\ab\,8\,pc\,\x\,8\,pc), and orientation is indicated. }
\label{finder}
\end{center}
\end{figure*}

\begin{figure}[!ht]
\resizebox{\hsize}{!}{\includegraphics{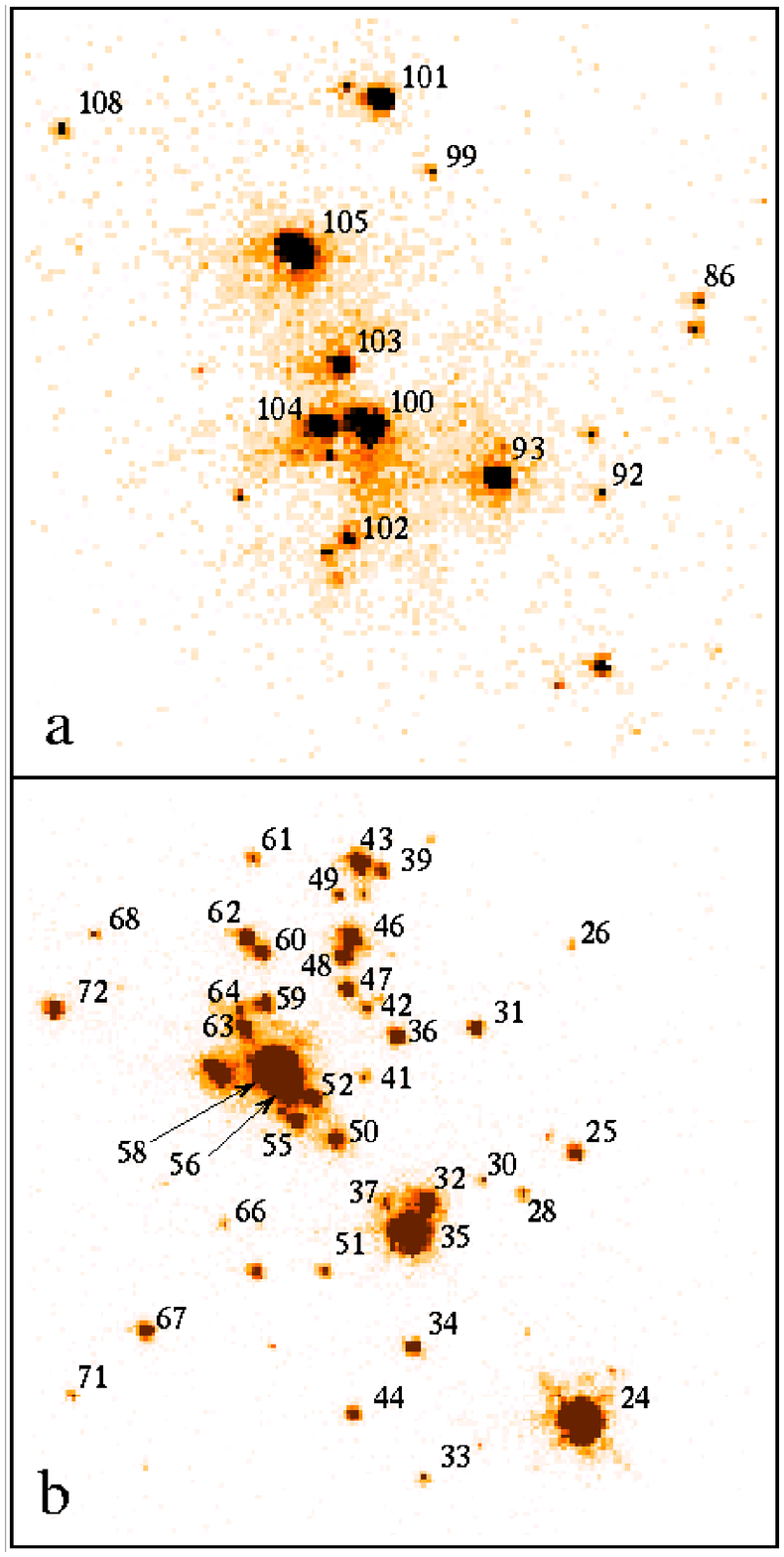}}
\caption{Details of the {\it HST}/WFPC2 image in the Str\"omgren $y$ filter
showing: a) the star cluster towards the compact \h2 region A2. The
central components are probably the exciting stars of A2, with star
\#105 ($y$\,=\,16.82 mag) being the main one.  The field size is
\ab\,8\frac\,\x\,8\frac\ (2\,pc\,\x\,2\,pc).  b) The central star
cluster lying between the compact \h2\, regions A1 and A2.  The field
size is \ab\,9\frac.3\,\x\,8\frac.5 (2.3\,pc\,\x\,2.1\,pc) and the
orientation is as in Fig.\,\ref{finder}.  }
\label{clusters}
\end{figure}

A crucial point in our data reduction was the sky subtraction. For
most isolated stars the sky level was estimated and subtracted
automatically using an annulus of 6--8 pixel width around each star.
However this could not be done for several stars located in the
central region of N160A due to their crowding. In those cases we
carefully examined the PSF size of each individual star ({\sc
fwhm}\,\ab\,2 pixels, corresponding to 0\frac.09 on the sky) and did
an appropriate sky subtraction using the mean of several nearby
off-star positions.  To convert into a magnitude scale we used zero
points in the Vegamag system, that is the system where Vega is set to
zero mag in Cousin broad-band filters.  The magnitudes measured were
corrected for geometrical distortion, finite aperture size (Holtzman
et al. \cite{holtz}), and charge transfer efficiency as recommended by
the {\it HST} Data Handbook. Our broad-band images reveal 110 stars
within the area covered by the PC. Most of them are also visible in
the true-color image (Fig.\,\ref{true}), and can be identified using
the finder charts presented in Figures\,\ref{finder} and
\ref{clusters}. In Table\,\ref{phot} we summarize the photometry for those
stars around in the PC2 field of view which are brighter than 19th
magnitude in the Str\"omgren $y$ filter, as we cannot provide accurate
colors for the fainter ones. The photometric errors estimated by {\it
daophot} are smaller than 0.01 mag for the brighter (14--15 mag)
stars, while they increase to $\sim$\,0.2 mag for 19 mag stars. \\

We note that the filter F547M is wider than the standard Str\"omgren
$y$ filter. To evaluate the presence of any systematic effects in our
photometry and color magnitude diagrams due to this difference in the
filters, we used the {\sc stsdas} package {\it synphot}.  Using
synthetic spectra of hot stars, with spectral types similar to those
found in \h2\, regions, we estimated the difference due to the {\it
HST} band-passes to be less than 0.002 mag, which is well within the
photometric errors. \\

\section{Results}

\subsection{Morphology}

Ground-based images show N160A as a very bright and rather elongated
emission nebula of size \ab\,35\frac\,\x\,25\frac, corresponding to
8.8\,\x\,6.3\,pc (Paper I). Three bright stars were identified within
the nebula along with the two outstanding bright compact \h2 ``blobs''
A1 and A2 separated by \ab\,20\frac\, (5 pc), 
However, those images were incapable of
resolving the structure of the blobs and the star cluster associated
with those three bright members 
seen as stars \#24, \#35, and \#58 plus their adjacent components in
Fig. 2 presented below. \\

A true color image of N160A, obtained with WFPC2 (Fig.\,\ref{true}),
displays a magnificent scene with evidence of high activity from
newborn massive stars: outstanding emission ridges sculpted by
powerful shocks and winds, arcs and filaments, several ``small'' dust
pillars protruding from hot gas, prominent dust concentrations
confining the ionized gas at the western side where a long undulating
ionization front is visible, etc.  More importantly, the {\it HST}
images for the first time resolve the HEBs revealing their
morphologies and stellar contents. The compact \h2\, region A1, which
is the brightest part of the whole N160A nebula, shows a ``tiny''
cavity or bubble, some 2\frac.3 across, carved by the strong wind of a
relatively bright star (\#22). The cavity has a remarkably thin edge
of $<$\,0\frac.2, but becomes thicker and more luminous in its
southeastern part. The central star appears to be offset towards that
direction. A1 is separated by a low brightness gap from the long
wavy-form bright ridge seen at \ab\,3\frac\, from star \#22. The other
HEB, A2, is 3\frac\, in diameter and displays a rather different
structure. It has a patchy appearance marked by the presence of
several thin absorption lanes, the main one situated towards its
central parts, and unlike A1, it contains several stars (See below
Sect. 3.4).\\

Our imaging of N160A also unveils at least three previously
unknown and even smaller ionized regions of size \ab\,1\frac\, located
in the vicinity of A2. They appear as reddish spots on the true
color image, and  are centered on stars \#108, \#29, and
\#72.  The reason for the color is either a cooler temperature of
their exciting stars or higher extinction due to the presence of
dust.  \\

\subsection{Extinction}

\begin{figure*}
\resizebox{\hsize}{!}{\includegraphics{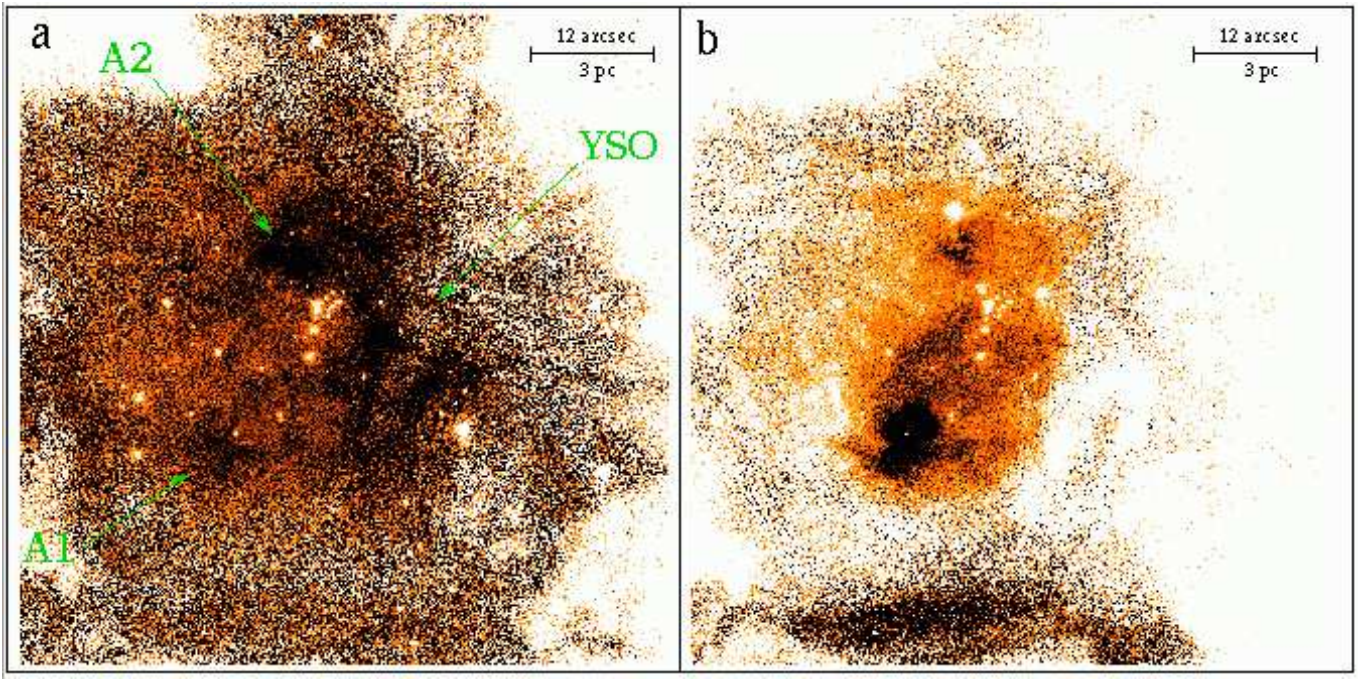}}
\caption{Line intensity ratios for the LMC compact nebula
N160A. Darker colors correspond to higher ratio values.  The field of
view and orientation are identical to Fig.\,\ref{true}. The white
spots are stars and can be identified using Figs.\,\ref{finder} and
\ref{clusters}.  {\bf a)} Balmer decrement \ha/\hb. Its mean value
over the diffuse component is \ab\,4.25 ($A_{V}$\,=\,1.1 mag), while
the ratio goes up to \ab\,4.5 ($A_{V}$\,=\,1.3 mag) and 4.9
($A_{V}$\,=\,1.5 mag) towards the compact blobs A1 and A2
respectively.  The position of the infrared young stellar object (see
text) is indicated.  {\bf b)} The [O\,{\sc iii}]$\lambda$5007/\hb\,
ratio. The mean value for the nebula outside the blobs is \ab\,4,
reaching its peak ($>$\,7) towards A1.}
\label{rapports}
\end{figure*}

A map of the \ha/\hb\, Balmer decrement is presented in
Fig.\,\ref{rapports}a and it further confirms that the \h2
region N160A is generally affected by a considerable amount of
interstellar dust. More importantly, it displays for the first time
the spatial distribution of the dust over the \h2 region.  It is
evident from this map that the dust is not uniformly distributed but
is rather ``patchy'' in nature.  In particular a remarkable quantity
of dust is concentrated behind the large western ionization front
where the \ha/\hb\, ratio has an average value of 4.76
($A_{V}$\,=\,1.4 mag) and peaks at as high as 6.33 ($A_{V}$\,=\,2.2
mag). The highest dust content though can be found further west of
this front, where the Balmer decrement has a mean value of 5.44
($A_{V}$\,=\,1.8 mag) reaching up to 6.87 ($A_{V}$\,=\,2.5 mag). The
average value of \ha/\hb\, towards the central regions of N160A is
4.25 ($A_{V}$\,=\,1.1 mag). \\

As one would expect, the two blobs lie in very dust rich areas of the
region. For A1 dust is mainly located behind the southern border of
the cavity, where the mean and maximum value of of the Balmer
decrement is 4.50 ($A_{V}$\,=\,1.3 mag), and 5.18 ($A_{V}$\,=\,1.7
mag) respectively. Interestingly, the northern compact \h2 region A2
is more affected by dust, covering a more extended area with a mean
Balmer decrement of 4.90 ($A_{V}$\,=\,1.5 mag) and attaining a maximum
value of 5.50 ($A_{V}$\,=\,1.8 mag). The extinction values reported
here revise the preliminary estimates given in Paper I.\\

The \ha/\hb\, map was also used to accurately correct the \hb\, flux
of the \h2 region for interstellar reddening.  The correction was
applied to the \hb\, image on a pixel by pixel basis using
straightforward mathematical operations.

\subsection{Nebular emission}

The [O\,{\sc iii}]$\lambda$5007/\hb\, intensity map displaying the
excitation of the \h2\, region N160A is presented in
Fig.\,\ref{rapports}b. A remarkable feature is the compact \h2\,
region A1 which stands out as the most excited part of N160A. The
bubble-shape nature of A1, apparent in Fig.\,\ref{true}, is also
obvious in lower contrast images of the \oiii/\hb\, ratio, confirming
the presence of a strong wind from the hot exciting star.  The mean
value of the ratio of the shell surrounding the cavity is \ab\,6.5,
with several localized peaks of $>$ 7. The map also reveals that the
curved ridge west of A1 is quite excited with a mean ratio of
\ab\,5.5. In contrast, the compact \h2 region A2 is less excited,
the \oiii/\hb\, ratio getting a mean value of \ab\,4.5 and never
exceeding \ab\,5.  Outside the compact regions A1 and A2, the ratio
has a mean value of \ab\,4, but it extends over the whole N160A. This
means that the O$^{++}$ ions occupy almost the same volume as H$^{+}$
necessitating the presence of several ionizing sources. \\

The total \hb\, flux of N160A was derived using the method described
in Section 3.2.  The corrected flux is
$F_{0}$(\hb)\,=\,2.20\,$\times$\,10$^{-10}$ erg cm$^{-2}$ s$^{-1}$
above 3$\sigma$ level accurate to 5\,\%.  Assuming that the \h2 region
is ionization-bounded, the corresponding Lyman continuum flux of N160A
is $N_{L}$\,=\,1.70\,$\times$\,10$^{50}$ photons s$^{-1}$.  Several
combinations of stars of various types can account for the observed
ionizing UV flux: two O3\,V and one O6\,V star, three O4\,V and one
O6\,V star, or even five O5\,V and one late O or one B0\,V star (Vacca
et al. \cite{vacca}, Schaerer \& de Koter \cite{sch}). However these
fluxes are probably underestimates since the \h2 region is more than
likely not completely ionization-bounded, due to the fact that it has
become open to the interstellar medium in the direction of the
observer. \\

In a similar manner one can estimate the de-reddened \hb\, fluxes of
both HEBs.  We find that $F_{0}$(\hb)\,=\,1.10\,$\times$\,10$^{-11}$
in A1, and 1.55\,$\times$\,10$^{-11}$ erg cm$^{-2}$ s$^{-1}$ in A2,
accurate to 20 and 30\% respectively.  Note that even though the
un-corrected images show the contrary, A2 is intrinsically more
luminous than A1 because it is more dusty.  The corresponding
estimated Lyman continuum photon fluxes are
$N_{L}$\,=\,8.50\,$\times$\,10$^{48}$\, photons s$^{-1}$for A1 and
$N_{L}$\,=\,1.20\,$\times$\,10$^{48}$\,photons s$^{-1}$ for A2.  A
single main sequence star of type earlier than O7.5--O8 can account
for the ionizing flux of A1, while in the case of A2 it should be at
least an O7--O7.5. \\

\subsection{Stellar content}

The {\it HST} images reveal some 110 stars brighter than $y$\,\ab\,21
mag across N160A. The brightest ones are grouped in a central cluster
extending over an area \ab\,9\frac\,\x\,8\frac\, between the two
compact \h2 regions A1 and A2. The brightest component of the
cluster, star \#58, has $y=14.56$ mag, and it is followed by stars
\#24, \#35, and \#56 with $y=14.70$, 15.08, and 15.23 mag respectively. The
region around star \#58 is rather crowded so it is likely that it
contains more stars which are not resolved with the PC2. \\

As we mentioned in Section 3.1, the images of blob A1 uncover
only one relatively bright star, \#22 with $y=15.66$ mag, lying inside
it. The star should be a massive O type because it has sculpted a
cavity and produced the highest \oiii/\hb\, ratio in the whole
region. The second blob A2, contains a dozen rather faint stars, the
brightest of which, \#105, has a magnitude $y=16.82$. The second
brightest star of A2 is \#100, with $y=17.22$ mag. \\

\begin{figure*}
\begin{center}
\resizebox{\hsize}{!}{\includegraphics{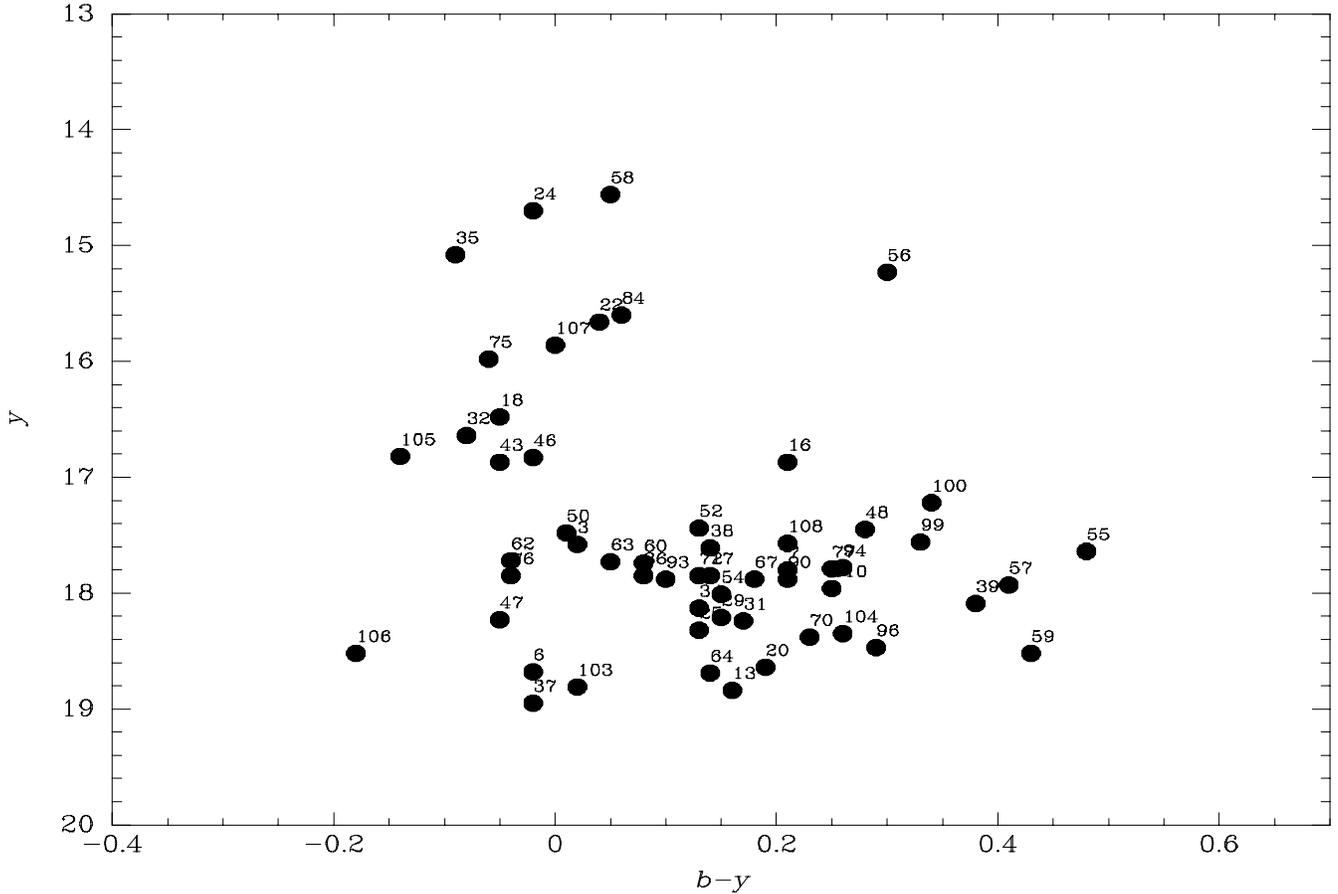}}
\caption{Color-magnitude diagram of the brightest stars 
(lower cutoff at $y=19$ mag) observed towards the \h2\, region N160A
based on WFPC2 imaging with the Str\"omgren filters $b$ (F467M) and
$y$ (F547M). The magnitudes are not corrected for reddening.}
\label{cm}
\end{center}
\end{figure*}

Using a cutoff at $y$\,=\,19 mag, we construct a color-magnitude
(C-M) diagram of $y$ versus $b -y$ (Fig.\,\ref{cm}) for the stars
observed across N160A. The diagram displays two principal
populations. A main sequence cluster in the interval $14.56 \leq y
\leq 19.00$ is centered  around $b - y$\,\ab\,$-0.05$ mag, while
there is a clear spread in colors due to the important and
inhomogeneous dust extinction towards the nebula, discussed in Section
3.2. The second stellar population, of mainly fainter stars, likely
consists of very reddened main sequence components and perhaps more
evolved field stars. It is quite possible that this latter population
is not physically associated with N160A and it may be along a line of
sight contamination to this young region. \\

One could try to estimate the luminosity of the brightest star of the
region (\#58), although in the absence of spectroscopic data this
would not be  very accurate. Using a reddening of $A_{V}$\,=\,1.1
mag corresponding to the mean value for the associated nebula (Section
3.2), and a distance modulus $m$\,--\,$M$\,=\,18.5 (e.g.  Kov\'acs
\cite{ko} and references therein), we find a visual absolute magnitude
$M_{V}=-5.04$ mag. Following the calibration of Vacca et
al. (\cite{vacca}) for Galactic stars, if the star is on the main
sequence, it would be an O6.5\,V, with a luminosity {\it log
L}\,=\,5.49\,$L_{\odot}$\, and a mass $M$\,=\,40\,\sm.\\

The other massive stars contributing to the ionization of the whole
\h2 region N160A are the brightest members of the central cluster,
i.e. \#58, \#24, \#35, and \#56. There are also some candidates lying
outside the cluster: \#84, \#107, \#75, and \#18.  We note that star
\#56 is offset red-wards ($b-y$\,\ab\,+0.30) in the C-M
diagram. However, its reddened color is  due to the presence of
dust, since the \ha/\hb\, ratio indicates a rather high extinction in
that direction, and moreover one of the ``small'' dust pillars lies 
there.  As a result, star \#58 is more than likely
intrinsically blue and one of the main exciting stars of the
region. \\

\section{Discussion}

The resolving power of {\it HST} makes it possible to see the details
of the interplay between the hot gas and prominent dust structures
present in this region.  In particular, the dust content, which is
quite high, increases westward where the remarkable ionization
front lies. The front must represent the interface with the molecular cloud
component N160-4 which appears to be adjacent to N160A
(Johansson et al. \cite{joh}). Its physical characteristics given by
the latter authors are: $V_{LSR}$\,=\,237.0 km\,$^{-1}$,
$\Delta$V\,=\,4.66 km\,s$^{-1}$, size 10.3 pc, CO luminosity
4.2\,\x\,10$^{3}$\,\,K\,km\,\,$s^{-1}$ pc$^{2}$, and virial mass
3.4\,\x\,10$^{4}$ \sm.  \\

Furthermore, the \ha/\hb\, map, which displays the variation of the
dust content across the nebula, shows rather high extinction
values. However, since the Balmer decrement is preferentially biased
towards emission from the outer regions of the nebula, we cannot probe
the deeply embedded IR sources, particularly the young stellar object
observed towards N160 (Epchtein et al. \cite{epchtein}, Henning et
al. \cite{henning}) for which the latter authors find a tremendous
extinction of $A_{V}$\,=\,60\,$\pm$\,10 mag.  Nonetheless, using the
reported coordinates of $\alpha$\,(2000)\,=\,$ 5^{h} 39^{m}43^{s}.8
\pm 0^{s}$.4, $\delta$\,(2000)\,=\,--69\deg\, 38\min\, 33\frac\,
$\pm$\,2\frac\, (Henning et al. \cite{henning}), we can precisely
locate it with respect to the \h2 region N160A. It lies behind the
ionization front in the absorption zone, or put loosely in the
``ghost's mouth'' (Figs.\,\ref{true} and \ref{rappors})! This
situation is very interesting for models studying the sequential star
formation (Elmegreen \& Lada \cite{el}). According to these models,
stars can form from the collapse of a narrow compressed shell between
the ionization front and the shock preceding it in the molecular
cloud. The YSO may therefore be a newborn star triggered by the
ionization front of N160A.\\

The two HEBs A1 and A2 represent very recent massive star formation
events in the N160A region, since all other massive stars have had
enough time to disrupt their natal material. Only the YSO can be
younger than them. A1, which is of higher excitation than A2, is
apparently powered by only one massive star (\#22) of type at least
O7.5--O8\,V. However, other undetected stars may be embedded in the dust
and gas.  The strong wind of star \#22 has carved the cavity or bubble
which can be easily seen in the {\it HST} images, and whose age can be
estimated. Assuming a typical mass loss rate of 6\,\x\,10$^{-5}$
\sm\,\,yr$^{-1}$ for the presumed central O star, a wind  velocity of
1000\,km\,s$^{-1}$, a gas density of 1500\,cm$^{-3}$ (Paper I), and an
observed radius of 1\frac.2 (0.3 pc) we find a lifetime of 2500 yr
from the classical equations governing the interaction of the stellar
wind and the interstellar medium (Weaver et al. \cite{weaver}, Dyson
\cite{dyson}). Since the estimated density is based on the [S\,{\sc
ii}] doublet, which is sensitive to the outer low density regions, we
may underestimate the real density. Should the density be a factor of
say 10 larger and the wind velocity a factor of 2 lower, the
corresponding age will be 8600 yr, that is very young in any case.  As
to A2, it may even be younger than A1 due to its heavy dust
content. We cannot confirm that all of the detected stars in A2 are
massive ones, but the high \oiii/\hb\, ratio and the measured \hb\,
flux indicate beyond doubt that massive stars are present. \\

Several of the N160A candidate stars (\#58, \#24, \#35, and \#56)
belong to the central cluster, while the rest (\#84, \#107, \#75, and
\#18) lie far from it. It is quite possible though that these
latter stars have been originally members of the central
cluster. Models studying formation of massive stars predict that they
should never form in isolation (Bonnell et al. \cite{bon}), and that
those found in isolation have been ejected from dense stellar clusters
(Leonard \& Duncan \cite{leonard}, Kroupa \cite{kroupa}). This
scenario is particularly attractive for very young massive star
regions like N160A where a dense central blue star cluster is present.
Escape velocities of newly formed stars can exceed 200\,km\,s$^{-1}$
(Leonard \& Duncan \cite{leonard}, Kroupa \cite{kroupa}), but the
largest fraction of them (23\%) is expected to have values of about
50\,km\,s$^{-1}$. Assuming such a velocity for star \#87, which has a
projected distance of \ab\,20\frac\, (5\,pc) from the cluster center,
it would take some 100\,000 years for this star to reach its present
position.  This estimate is of course a lower limit, since we use
projected and not the real distances.  Inversely, one can calculate
the minimum velocities at which the stars could have reached their
current locations. Assuming a lifetime of 3\,Myr, one gets a lower
ejection velocity of \ab\,2 km\,s$^{-1}$ for \#87.  One might wonder
whether the observed star density of the central cluster is
sufficiently high for the dynamical ejection mechanism to
work. However, Leonard \& Duncan (\cite{LD88}) have shown that
binary-binary collisions required to produce high velocity escapees
occur in low density clusters, even though simple estimates suggest
that such interactions are unlikely. Furthermore, the ejection of the
stars must have happened during an earlier evolutionary stage when the
cluster was most probably more compact than today (Portegies Zwart et
al.\,\cite{port99}).  \\

The population of red stars present in the color-magnitude diagram is
expected to actually be on the main sequence but affected by the dust
in this young region. However, with the present data we cannot exclude
the presence of evolved and/or lower mass pre-main sequence stars. If
evolved stars are really there, this would imply that star formation
has occurred in successive waves in this part of the LMC. Whether or
not lower mass stars are indeed present in this region is very
interesting as it could be a test for the so-called bi-modal star
formation concept predicting that high mass stars form without their
low mass counterparts (G\"usten \& Mezger \cite {gusten}, Zinnecker et
al. \cite{zin}).  The fact that HEBs in general, and N160A in
particular, are extremely young provides an unparallel opportunity for
verifying these models. \\

\begin{acknowledgements} 
We would like to thank the referee Dr. Joel Wm. Parker for his
helpful comments which improved the manuscript. VC would like to
acknowledge the financial support for this work provided by NASA
through grant number GO-8247 from the STScI, which is operated by the
Association of Universities for Research in Astronomy, Inc., under
NASA contract NAS\,5-26555.
\end{acknowledgements}

{}

\begin{table*}[t]	
\caption[]{{\it HST} Photometry of the brightest stars towards N160A}	
\label{phot}	
\begin{flushleft}
\begin{tabular}{cccccccc} 	
\hline	
  Star & RA(J2000) & Dec(J2000) & F300W & F410M & F467M & F547M &
  Color\\ 
   & & & Wide $U$ & Str\"omgren $v$ & Str\"omgren $b$ &
  Str\"omgren $y$ & $b-y$ \\
\hline	
  1 & 5:39:41.1 & -69:38:35.0 & --    & --    & --    & --    & -- \\
  2 & 5:39:41.0 & -69:38:49.0 & 18.60 & --    & --    & --    & --\\
  3 & 5:39:41.0 & -69:38:49.2 & 16.43 & 17.61 & 17.60 & 17.58 & 0.02\\
  4 & 5:39:41.3 & -69:38:36.9 & 18.28 & 19.27 & 19.22 & 19.02 & 0.20\\
  5 & 5:39:41.5 & -69:38:31.8 & 18.53 & 19.94 & 19.30 & --    & -- \\
  6 & 5:39:41.4 & -69:38:33.7 & 17.32 & 18.77 & 18.66 & 18.68 & -0.02\\
  7 & 5:39:41.6 & -69:38:30.3 & 16.44 & 17.96 & 18.01 & 17.80 & 0.21\\
  8 & 5:39:41.5 & -69:38:38.9 & --    & --    & --    & --    & --\\
  9 & 5:39:41.7 & -69:38:33.8 & 18.69 & --    & --    & --    & --\\
 10 & 5:39:42.0 & -69:38:32.6 & 18.39 & --    & --    & --    & --\\
 11 & 5:39:42.1 & -69:38:37.3 & --    & --    & --    & --    & --\\
 12 & 5:39:42.6 & -69:38:41.0 & 17.67 & 19.53 & 19.35 & 18.94 & 0.41\\
 13 & 5:39:42.3 & -69:38:56.7 & 17.40 & 18.92 & 19.01 & 18.84 & 0.17\\
 14 & 5:39:42.5 & -69:38:54.2 & 18.18 & 19.88 & 19.50 & 19.55 & -0.05\\
 15 & 5:39:42.8 & -69:38:42.9 & 18.12 & 19.83 & 19.79 & 19.58 & 0.21\\
 16 & 5:39:43.0 & -69:38:38.6 & 15.89 & 17.04 & 17.08 & 16.87 & 0.21\\
 17 & 5:39:43.1 & -69:38:48.7 & 16.98 & 18.10 & 17.99 & 17.85 & 0.14\\
 18 & 5:39:43.2 & -69:38:48.5 & 15.23 & 16.45 & 16.43 & 16.48 & -0.05\\
 19 & 5:39:43.4 & -69:38:44.9 & --    & --    & --    & --    & -- \\
 20 & 5:39:43.4 & -69:38:41.6 & 17.38 & 18.79 & 18.83 & 18.64 & 0.19\\
 21 & 5:39:43.2 & -69:38:58.2 & 18.56 & --    & --    & --    & 0.97\\
 22 & 5:39:43.4 & -69:38:53.6 & 14.27 & 15.55 & 15.70 & 15.66 & 0.04\\
 23 & 5:39:44.0 & -69:38:32.8 & 18.24 & 19.57 & 19.38 & 19.04 & 0.34\\
 24 & 5:39:43.9 & -69:38:42.4 & 13.22 & 14.51 & 14.68 & 14.70 & -0.02\\
 25 & 5:39:44.0 & -69:38:39.4 & 17.25 & 18.29 & 18.45 & 18.32 & 0.13\\
 26 & 5:39:44.0 & -69:38:37.1 & 18.55 & --    & --    & --    & --\\
 27 & 5:39:44.1 & -69:38:35.8 & --    & --    & --    & --    & --\\
 28 & 5:39:44.1 & -69:38:40.0 & 18.68 & 19.46 & 19.80 & 19.49 & 0.31\\
 29 & 5:39:44.3 & -69:38:32.9 & 16.97 & 18.32 & 18.36 & 18.21 & 0.15\\
 30 & 5:39:44.2 & -69:38:39.9 & --    & --    & --    & --    & -- \\
 31 & 5:39:44.2 & -69:38:38.2 & 16.93 & 18.28 & 18.41 & 18.24 & 0.17\\
 32 & 5:39:44.3 & -69:38:40.2 & 15.28 & 16.46 & 16.56 & 16.64 & -0.08\\
 33 & 5:39:44.2 & -69:38:43.2 & 19.58 & --    & --    & 19.71 & -- \\
 34 & 5:39:44.3 & -69:38:41.8 & 16.95 & 18.15 & 18.26 & 18.13 & 0.13\\
 35 & 5:39:44.3 & -69:38:40.5 & 13.67 & 14.88 & 14.99 & 15.08 & -0.09\\
 36 & 5:39:44.4 & -69:38:38.3 & 16.59 & 17.73 & 17.93 & 17.85 & 0.08\\
 37 & 5:39:44.4 & -69:38:40.2 & 18.10 & 18.98 & 18.93 & 18.95 & -0.02\\
 38 & 5:39:44.3 & -69:38:47.4 & 15.90 & 17.58 & 17.75 & 17.61 & 0.14\\
 39 & 5:39:44.5 & -69:38:36.5 & 17.50 & 18.65 & 18.47 & 18.09 & 0.38\\
 40 & 5:39:44.5 & -69:38:32.7 & 18.77 & --    & --    & --    & --\\
 41 & 5:39:44.5 & -69:38:38.8 & --    & 19.84 & 19.88 & --    & --\\
 42 & 5:39:44.5 & -69:38:38.1 & 18.73 & 19.77 & 19.16 & 19.27 & -0.11\\
 43 & 5:39:44.5 & -69:38:36.5 & 15.71 & 16.94 & 16.82 & 16.87 & -0.05\\
 44 & 5:39:44.4 & -69:38:42.6 & 18.03 & 18.62 & 18.93 & 18.66 & 0.27\\
 45 & 5:39:44.2 & -69:38:53.6 & 18.23 & --    & --    & 19.58 & --\\
 46 & 5:39:44.5 & -69:38:37.3 & 15.68 & 16.84 & 16.81 & 16.83 & -0.02\\
 47 & 5:39:44.5 & -69:38:37.9 & 16.99 & 18.28 & 18.18 & 18.23 & -0.05\\
 48 & 5:39:44.5 & -69:38:37.5 & 16.65 & 17.67 & 17.73 & 17.45 & 0.28\\
 49 & 5:39:44.6 & -69:38:36.9 & 19.65 & --    & 19.52 & 19.85 & -0.33\\
 50 & 5:39:44.5 & -69:38:39.6 & 16.29 & 17.45 & 17.49 & 17.48 & 0.01\\
 51 & 5:39:44.5 & -69:38:41.0 & 18.52 & 19.56 & 19.74 & 19.51 & 0.23\\
 52 & 5:39:44.6 & -69:38:39.2 & 16.41 & 17.70 & 17.57 & 17.44 & 0.13\\
 53 & 5:39:44.6 & -69:38:35.6 & 18.51 & --    & --    & --    & --\\
 54 & 5:39:44.3 & -69:38:56.4 & 16.54 & 17.94 & 18.16 & 18.01 & 0.15\\
 55 & 5:39:44.6 & -69:38:39.4 & 16.90 & 17.97 & 18.12 & 17.64 & 0.48\\
 56 & 5:39:44.6 & -69:38:39.0 & 14.05 & 15.56 & 15.53 & 15.23 & 0.30\\
 57 & 5:39:44.6 & -69:38:39.3 & 17.48 & 18.11 & 18.34 & 17.93 & 0.41\\
 58 & 5:39:44.7 & -69:38:38.8 & 13.26 & 14.48 & 14.61 & 14.56 & 0.05\\
 59 & 5:39:44.7 & -69:38:38.2 & 18.14 & 18.89 & 18.95 & 18.52 & 0.43\\
 60 & 5:39:44.7 & -69:38:37.6 & 16.61 & 17.87 & 17.82 & 17.74 & 0.08\\
\end{tabular}
\end{flushleft}
\end{table*}

\begin{table*}[t]	
\begin{flushleft}
\begin{tabular}{cccccccc} 	
\hline	
  Star & RA(J2000) & Dec(J2000) & F300W & F410M & F467M & F547M &
  Color\\ 
   & & & Wide $U$ & Str\"omgren $v$ & Str\"omgren $b$ &
  Str\"omgren $y$ & $b-y$ \\
\hline	
 61 & 5:39:44.8 & -69:38:36.5 & 17.84 & 19.19 & 19.36 & 19.03 & 0.33\\
 62 & 5:39:44.8 & -69:38:37.4 & 16.51 & 17.57 & 17.68 & 17.72 & -0.04\\
 63 & 5:39:44.7 & -69:38:38.4 & 16.79 & 17.78 & 17.78 & 17.73 & 0.05\\
 64 & 5:39:44.7 & -69:38:38.2 & 18.95 & 18.56 & 18.83 & 18.69 & 0.14\\
 65 & 5:39:44.6 & -69:38:48.1 & 18.69 & --    & --    & --    & -- \\
 66 & 5:39:44.7 & -69:38:40.6 & --    & --    & --    & 19.81 & -- \\
 67 & 5:39:44.9 & -69:38:41.9 & 16.70 & 17.89 & 18.06 & 17.88 & 0.18\\
 68 & 5:39:45.1 & -69:38:37.6 & 18.55 & 19.90 & 19.84 & 19.71 & 0.13\\
 69 & 5:39:45.2 & -69:38:34.0 & 18.48 & --    & --    & 19.84 & --\\
 70 & 5:39:45.2 & -69:38:33.6 & 17.08 & 18.41 & 18.61 & 18.38 & 0.23\\
 71 & 5:39:45.0 & -69:38:42.7 & 19.43 & --    & --    & 19.83 & --\\
 72 & 5:39:45.2 & -69:38:38.4 & 16.90 & 17.98 & 17.98 & 17.85 & 0.13\\
 73 & 5:39:45.0 & -69:38:47.2 & 18.27 & --    & --    & 19.28 & --\\
 74 & 5:39:45.4 & -69:38:31.6 & 19.00 & 19.36 & 19.27 & 18.81 & 0.46\\
 75 & 5:39:45.0 & -69:38:50.4 & 14.51 & 15.92 & 15.92 & 15.98 & -0.06\\
 76 & 5:39:45.2 & -69:38:40.5 & 16.64 & 17.85 & 17.81 & 17.85 & -0.04\\
 77 & 5:39:45.4 & -69:38:33.6 & 16.44 & 17.85 & 18.04 & 17.79 & 0.25\\
 78 & 5:39:45.4 & -69:38:32.2 & 18.77 & --    & --    & 19.59 & --\\
 79 & 5:39:45.3 & -69:38:38.7 & 18.98 & --    & --    & 19.98 & --\\
 80 & 5:39:45.2 & -69:38:47.4 & 18.80 & --    & --    & --    & --\\
 81 & 5:39:45.0 & -69:38:58.0 & 17.88 & 19.30 & 19.36 & 19.00 & 0.36\\
 82 & 5:39:45.4 & -69:38:41.4 & 18.79 & 19.98 & --    & 19.86 & -- \\
 83 & 5:39:45.4 & -69:38:46.7 & 17.96 & 19.95 & 19.67 & 19.52 & 0.15\\
 84 & 5:39:45.1 & -69:39:00.1 & 14.06 & 15.47 & 15.66 & 15.60 & 0.06\\
 85 & 5:39:45.2 & -69:38:55.6 & 18.12 & 19.89 & --    & 19.51 & --\\
 86 & 5:39:45.6 & -69:38:36.8 & 19.62 & --    & --    & --    & 1.15\\
 87 & 5:39:45.2 & -69:39:00.3 & 18.98 & 19.62 & 19.58 & 19.48 & 0.10\\
 88 & 5:39:45.5 & -69:38:52.2 & 17.92 & 19.56 & 19.61 & 19.14 & 0.47\\
 89 & 5:39:45.4 & -69:38:56.1 & 17.57 & 19.18 & 19.15 & 18.98 & 0.17\\
 90 & 5:39:45.7 & -69:38:41.6 & 16.77 & 17.91 & 18.09 & 17.88 & 0.21\\
 91 & 5:39:45.3 & -69:39:02.2 & 18.47 & 19.57 & --    & 19.36 & --\\
 92 & 5:39:45.8 & -69:38:38.4 & 19.49 & --    & --    & --    & --\\
 93 & 5:39:45.9 & -69:38:38.4 & 16.71 & 17.87 & 17.98 & 17.88 & 0.10\\
 94 & 5:39:45.5 & -69:39:01.0 & 16.52 & 17.92 & 18.04 & 17.78 & 0.26\\
 95 & 5:39:45.8 & -69:38:43.4 & 19.95 & --    & --    & --    & --\\
 96 & 5:39:45.7 & -69:38:51.2 & 17.32 & 18.59 & 18.76 & 18.47 & 0.29\\
 97 & 5:39:45.9 & -69:38:43.4 & 19.00 & --    & --    & --    & --\\
 98 & 5:39:45.7 & -69:38:52.2 & 18.18 & --    & --    & 19.88 & -- \\
 99 & 5:39:46.1 & -69:38:35.5 & 16.43 & 17.68 & 17.89 & 17.56 & 0.33\\
100 & 5:39:46.1 & -69:38:38.0 & 16.30 & 17.62 & 17.56 & 17.22 & 0.34\\
101 & 5:39:46.2 & -69:38:34.8 & 18.59 & 19.39 & 19.72 & 19.27 & 0.45\\
102 & 5:39:46.1 & -69:38:38.9 & 18.52 & 19.67 & 19.65 & 19.82 & -0.17\\
103 & 5:39:46.2 & -69:38:37.6 & 19.53 & 19.30 & 18.83 & 18.81 & 0.02\\
104 & 5:39:46.2 & -69:38:38.1 & 17.21 & 18.48 & 18.61 & 18.35 & 0.26\\
105 & 5:39:46.2 & -69:38:36.7 & 15.37 & 16.76 & 16.68 & 16.82 & -0.14\\
106 & 5:39:45.9 & -69:39:02.8 & 17.13 & 18.45 & 18.34 & 18.52 & -0.18\\
107 & 5:39:46.4 & -69:38:52.1 & 14.41 & 15.75 & 15.86 & 15.86 & 0.00\\
108 & 5:39:46.7 & -69:38:36.4 & 16.29 & 17.54 & 17.78 & 17.57 & 0.21\\
109 & 5:39:46.7 & -69:38:48.9 & 18.19 & --    & 19.03 & 19.17 & -0.14\\
110 & 5:39:46.8 & -69:38:44.4 & 16.60 & 18.10 & 18.21 & 17.96 & 0.25 \\ 
\end{tabular}
\end{flushleft}
\end{table*}


\begin{thebibliography}{}

\bibitem[1998]{bon}
   Bonnell I.A., Bate M.R., Zinnecker H., 1998, MNRAS 298, 93
\bibitem[1997]{brooks}
   Brooks K.J., Whiteoak J.B., 1997, MNRAS 291, 395
\bibitem[1981]{caswell1}
   Caswell J.L., Haynes R.F., 1981, MNRAS 194, 33P
\bibitem[1995]{caswell2}
   Caswell J.L., 1995, MNRAS 272, L31
\bibitem[1976]{dem}
   Davies, R.D. Elliott K.H., Meaburn J., 1976, MNRAS 81, 89
\bibitem[1978]{dyson}
   Dyson J.E., 1978, A\&A 62, 269
\bibitem[1977]{el}
   Elmegreen B.G., Lada C.J., 1977, ApJ 214, 725
\bibitem[1984]{epchtein}
   Epchtein N., Braz M.A., S\`evre F., 1984, A\&A 140, 67
\bibitem[1985]{gardener}
   Gardener F.F., Whiteoak J.B., 1985, MNRAS 215, 103
\bibitem[1982]{gusten}
   G\"usten R., Mezger P.G., 1982, Vistas in Astronomy 26, 159
\bibitem[1956]{hen}
   Henize K.G., 1956, ApJS 2, 315
\bibitem[1998]{henning}
   Henning Th., Klein R., Chan S.J., Fitzpatrick E.L., et al. 1998, 
      A\&A 338, L51
\bibitem[1986]{hey86}
   Heydari-Malayeri M., Testor G., 1986, A\&A 162, 180  (Paper I)
\bibitem[2001]{hey01} 
Heydari-Malayeri M., Charmandaris V., Deharveng L., Rosa M.R., Schaerer D., 
      Zinnecker H., 2001, A\&A 372, 495 
\bibitem[1995]{holtz}
   Holtzman J., Hester J.J, Casertano S., et al., 1995, PASP 107, 156
\bibitem[1998]{joh}
   Johansson L.B.E., Greve A., Booth R.S. et al., 1998, A\&A 331, 857
\bibitem[1986]{jon}
   Jones T.J., Hyland A.R., Straw S., et al., 1986, MNRAS 219, 603
\bibitem[2000]{ko}
   Kov\'acs G., 2000, A\&A 363, L1
\bibitem[1995]{kroupa}
   Kroupa P., 1995, MNRAS 277, 1522   
\bibitem[1988]{LD88}
   Leonard, P.J.T., Duncan M.J., 1988, AJ 96, 222  
\bibitem[1990]{leonard}
   Leonard, P. J. T., Duncan M.J., 1990, AJ 99, 608
\bibitem[1974]{lucke}
   Lucke P.B., 1974, ApJS 28, 73 
\bibitem[1970]{lh}
   Lucke B.P., Hodge P.W., 1970, AJ 75, 171
\bibitem[1966]{mc}
   McGee R.X., Milton J.A., 1966, Aust. J. Phys. 19, 343
\bibitem[1999]{port99}
   Portegies Zwart S.E., Makino J., McMillan S.L.W., Hut P., 
   1999, A\&A 348, 117
\bibitem[1978]{rk}
   Relyea L.J., Kurucz R.L., 1978, ApJS 37, 45
\bibitem[1997]{sch}
   Schaerer D., de Koter A., 1997, A\&A 322, 598
\bibitem[1996]{vacca}
   Vacca W.D., Garmany C.D., Shull J.M., 1996, ApJ 460, 914
\bibitem[1977]{weaver}
   Weaver R., McCray R., Castor J., et al., 1977, ApJ 218, 377
\bibitem[1983]{w83}
   Whiteoak J.B., Wellington K.J., Jauncey D.L., et al., 1983, MNRAS 205, 275
\bibitem[1986]{w86}
   Whiteoak J.B., Gardener F.F., 1986, MNRAS 222, 513
\bibitem[1993]{zin}
Zinnecker H., McCaughrean M.J., Wilking B., 1993, in ``Protostars and 
   Planets III'', eds. E.H. Levy \& J.I. Lumine, Univ. of Arizona Press, 
   Tuscon, p. 429
\end{thebibliography}
\end{document}